# Hubbard Model on Decorated Lattices

C. D. Batista* and B. S. Shastry*,†,‡

*Theoretical Division, Los Alamos National Laboratory, Los Alamos, NM 87545
†Physics Department, Indian Institute of Science, Bangalore, 560012, India

We introduce a family of lattices for which the Hubbard model and its natural extensions can be quasi-exactly solved, i.e. solved for the ground and low energy states. In particular, we show rigorously that the ground state of the Hubbard model with off-site Coulomb repulsions on a decorated Kagomè lattice is an ordered array of local currents. The low energy theory describing this chiral state is an $S = \frac{1}{2}$ $XY$ model, where each spin degree of freedom represents the two possible chiralities of each local current.

Exactly solvable manybody problems have played an important role in condensed matter physics. Several solvable models are known in one dimension. The list in two and higher dimensions, while much smaller, is growing. A particularly fruitful idea for constructing new solvable models of interacting particles has emerged from the work of Refs. [1, 2]. This idea, building on several examples, has been formulated in terms of "superstable" ground states, where clusters of particles have local ground states that survive, or extend easily on connecting into a lattice. The Shastry-Sutherland S=1/2 Heisenberg model system is a rich example since the lattice is a natural one, theoretically in that there are no crossed bonds, and practically in its realization in the compound $SrCu_2(BO_3)_2$ [3].

So far this idea has been applied predominantly to spin systems. In this work we present what seems to be the first application of this idea to fermionic systems. Our main new result is a generalized Hubbard model in two dimensions on a decorated Kagomè lattice. The model has triangular units of circulating currents as superstable objects that further interact to give interesting quantum ordered states. We show rigorously that the ground state of this model has long range order for certain fillings in two dimensions. This state is an array of local currents with two possible orientations that live on the up and down triangles that comprise the Kagomè lattice. This internal degree of freedom localized in each triangle can be described as a two level system ($S = 1/2$ pseudo-spin) which represents two possible directions of circulating electronic current, or two chiralities. Similar solutions are currently popular in connection with theories of high $T_c$ superconductivity [4, 5] and heavy fermions [6]. A hidden order parameter which breaks the time-reversal symmetry has been proposed to exist underneath the superconducting dome in the phase diagram of the cuprates and to characterize the mysterious phase observed in $URu_2Si_2$ below $17°K$ [8]. The model is solvable for any filling $\rho \leq \rho_L$, displays long ranged order of a hidden sort and in fact maps onto a $S = \frac{1}{2}$ $XY$ model that is known to have Onsager Penrose LRO in two dimensions rigorously [9].

An important consequence of this simple and rigorous solution is the emergence of a new microscopic mechanism for the generation of an ordered chiral state. The geometrical frustration of the considered lattices together with the on-site Coulomb repulsion generate a state with non-zero local currents. The effective interaction between these currents is provided by the intersite Coulomb repulsion. In this way, the interplay between frustration and strong correlations gives rise to new types of orderings.

The main idea described in this paper can be applied to the family of lattices represented in Fig. 1. This family is a bipartite structure (B and C), with an arbitrary coordination number $z$ ($z = 2$ in Fig. 1a and $z = 3$ in Fig. 1b). The block B (similar in spirit to that in Ref([10])) represents a cell that will be called *basis* and the the block C represents the other non-equivalent cell which will be called *connector*. The blocks C are fully connected with its nearest neighbor (n.n.) blocks B; i.e., each site in a block C is connected to each of the sites in the n. n. block B. The sites in a block B have the same internal coordination number $z_B$, i.e. each site in B is connected to $z_B$ sites in the same block. Notice that the blocks of each type need not all be equivalent. Using the general structure indicated in Fig. 1, we can generate a family of lattices for which the low energy spectra of well known fermionic models (like the Hubbard, the $t-J$ or the interacting spin-less fermion Hamiltonian) can be exactly mapped into simple spin models.

For this family of lattices, we will consider an extended Hubbard model which includes Coulomb repulsions between different sites:

$$H = t \sum_{\langle \mathbf{r},\mathbf{r}'\rangle \in B,\sigma} (c^\dagger_{\mathbf{r}\sigma} c_{\mathbf{r}'\sigma} + c^\dagger_{\mathbf{r}'\sigma} c_{\mathbf{r}\sigma}) + U \sum_{\mathbf{r}} n_{\mathbf{r}\uparrow} n_{\mathbf{r}\downarrow} \\ + t' \sum_{\langle \mathbf{r} \in B,\mathbf{r}' \in C\rangle,\sigma} (c^\dagger_{\mathbf{r}\sigma} c_{\mathbf{r}'\sigma} + c^\dagger_{\mathbf{r}'\sigma} c_{\mathbf{r}\sigma}) + \sum_{\mathbf{r},\mathbf{r}',\sigma} V_{\mathbf{r},\mathbf{r}'} n_{\mathbf{r}} n_{\mathbf{r}'}, \quad (1)$$

where $\langle \mathbf{r}, \mathbf{r}' \rangle \in B$ indicates that $\mathbf{r}$ and $\mathbf{r}'$ are nearest neighbor sites which belong to the same block B, and $\langle \mathbf{r} \in B, \mathbf{r}' \in C \rangle$ indicates that B and C are nearest neighbor blocks. We will consider that the repulsive interaction $V_{\mathbf{r},\mathbf{r}'}$ is extended up to third n. n. sites ($V_1$, $V_2$ and $V_3$).

In Fig. 1a and 1b, we show two possible realizations of the lattice structure described above for one and two dimensions (decorated Kagomè lattice) respectively. The first one has been recently considered by Rojas *et al* to solve a Heisenberg model [7]. In these examples, each block B is a triangle of three sites (circles in Fig. 1) and the connectors C are single sites in between two triangles (squares in Fig. 1). In the two dimensional case (Fig. 1b), there are two different classes of blocks B. The different terms of $H$ acting on this lattice are also shown in Fig. 1a.

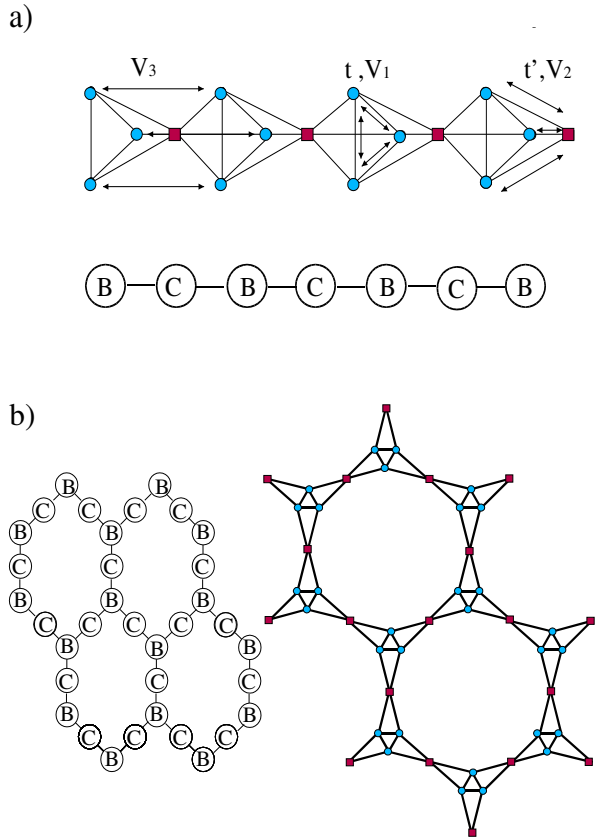

FIG. 1: a) 1D lattice made up of corner sharing "double-tetrahedra" aligned in a linear chain and the general scheme for this family of 1D lattices. The squares are the connectors (C) and the circles belong to the basis blocks (B). b) A possible general scheme for 2D lattices and a particular case of this family: the decorated Kagomè lattice.

*One Dimensional Lattice.* This lattice may be imagined as being made up of corner sharing "double-tetrahedra" aligned in a linear chain. The unit cell of the one dimensional (1D) lattice contains four sites: one block B or triangle and one connector (see Fig. 2b). For this reason, the noninteracting part of $H$, $H_0$, gives rise to four different bands. To analyze the structure of these bands we will first consider the eigenstates of $H_0$ acting on an isolated block B. Since B is a triangle of three sites connected by the hopping term $t$, the three quasi-particle operators and eigenenergies of $H_0$ restricted to B are:

$$\phi^\dagger_{\sigma 0} = \frac{1}{\sqrt{3}}(c^\dagger_{1\sigma} + c^\dagger_{2\sigma} + c^\dagger_{3\sigma})  \quad E_0 = 2t$$

$$\phi^\dagger_{\sigma +} = \frac{1}{\sqrt{3}}(c^\dagger_{1\sigma} + \omega c^\dagger_{2\sigma} + \omega^2 c^\dagger_{3\sigma}) \quad E_+ = -t$$

$$\phi^\dagger_{\sigma -} = \frac{1}{\sqrt{3}}(c^\dagger_{1\sigma} + \omega^2 c^\dagger_{2\sigma} + \omega c^\dagger_{3\sigma}) \quad E_- = -t, \quad (2)$$

where $\omega = e^{i\frac{2\pi}{3}}$ and the numbers 1, 2, and 3 denote the three different sites of the triangle. $\phi^\dagger_{\sigma 0}$ creates a local state with zero current, while $\phi^\dagger_{\sigma +}$ and $\phi^\dagger_{\sigma -}$ create local states with positive and negative chiralities respectively (see Fig. 2a). When the considered block B is reinserted in the lattice, the effective hybridization between a given $\phi^\dagger_{\eta\sigma}$ ($\eta$ is the chiral index) orbital and its nearest neighbor connector is: $t' \sum_{j=1}^{3} \langle 0|c_{j,\sigma} \phi^\dagger_{\eta\sigma}|0\rangle$. Therefore, $\phi^\dagger_{\mathbf{i}\sigma+}$ and $\phi^\dagger_{\mathbf{i}\sigma-}$ are still quasi-particle operators of $H_0$ because they do not hybridize with the connectors (the index $\mathbf{i}$ denotes the position of the triangle or block B). This means that $H_0$ contains two flat bands with energies $E^\pm(\mathbf{k}) = -t$ (the supra index denotes the chirality) for which $\phi^\dagger_{\mathbf{i}\sigma+}$ and $\phi^\dagger_{\mathbf{i}\sigma-}$ create the corresponding Wannier orbitals. The other two bands are generated by the hybridization between the orbitals created by $\phi^\dagger_{\mathbf{i}\sigma 0}$ and the connectors. Since the effective hopping between these two orbitals is $\sqrt{3}t'$, the corresponding dispersion relations are: $E^0_1(\mathbf{k}) = t - \sqrt{t^2 + 3t'^2}$ and $E^0_2(\mathbf{k}) = t + \sqrt{t^2 + 3t'^2}$.

If $t > 0$ and $t \gg |t'|$, the energy of the two degenerate flat bands is much lower than the energy of both dispersive bands. Note that this election for the sign of the hopping integral $t$ is realistic if we consider that $c^\dagger_{\mathbf{r}\sigma}$ and $c_{\mathbf{r}\sigma}$ are creation and anihilation operators for holes. The solution that we describe below is then valid for systems with low concentrations of holes ($0 \le \rho \le \frac{1}{8}$). In this situation, the lowest energy subspace of $H(V_3 = 0)$ can be exactly solved for any concentration $\rho \le \frac{1}{8}$. The ground states of $H(V_3 = 0)$ for a fixed concentration $\rho \le \frac{1}{8}$ are:

$$\phi^\dagger_{\mathbf{i}_1 \eta_1 \sigma_1} \ldots \phi^\dagger_{\mathbf{i}_\nu \eta_\nu \sigma_\nu} \ldots \phi^\dagger_{\mathbf{i}_N \eta_N \sigma_N} |0\rangle \quad (3)$$

where $N$ is the number of particles, $\eta_\nu$ is the chirality flavor of the particle $\nu$ and the block indices $\mathbf{i}_\nu$ are all different; i.e., there is no more than one quasi-particle per block B. This condition implies that the concentration of holes $\rho$ cannot be larger than $1/8$ $\rho \le \frac{1}{8}$. Notice that the index $\eta$ can denote any direction in the $SU(2)$ space generated by the linear combinations of the two possible chiralities + and −. The on-site and the nearest-neighbor Coulomb repulsions, $U$ and $V_1$, prevent the double occupancy on each triangle. In other words, the mean value of the $U$ and the $V_1$ terms is zero for the wave functions [3]. In addition, the mean value of the $V_2$ term is also zero because the connectors are empty. It is clear then that the wave functions [3] are the ground states of $H$ if $V_3 = 0$ since they minimize the mean value of each term in $H(V_3 = 0)$ when the concentration $\rho$ is fixed. In general there are three different sources for the massive degeneracy of the ground state: the spin orientation, the chiral orientation and the choice of the triangles which are occupied (charge degrees of freedom) if $\rho < \frac{1}{8}$.

It is clear that the local chirality and spin on each triangle are conserved quantities when the Hamiltonian $H(V_3 = 0)$ is restricted to the subspace generated by the states [3]. This conservation is related to an $SU(2) \times SU(2)$ local gauge symmetry which is dynamically generated at low energies. In addition, the local charge conservation on each triangle is associated to a local $U(1)$ invariance. Therefore, the total gauge symmetry group at low energies is $U(1) \times SU(2) \times SU(2)$. This emerging symmetry is a consequence of the particular geometry of the considered lattices and allows us to get the exact ground states for $\rho \le \frac{1}{8}$.

In particular, for $\rho = \frac{1}{8}$, there is one $\phi$ quasi-particle per block and the remaining degeneracy only comes from the two possible chiralities and spin orientations. The chiral degeneracy is removed when the inter-site Coulomb repulsion, $V_3$, is included.

To analyze the effect of $V_3$, we will replace the chiral index $\eta$ by a pseudo-spin flavor $\tau$ using the following convention [13](see Fig. 2a):

$$\begin{aligned}
\tau_{\mathbf{i}}^x &= \frac{1}{2}\sum_\sigma (\phi_{\mathbf{i}\sigma+}^\dagger \phi_{\mathbf{i}\sigma-} + \phi_{\mathbf{i}\sigma-}^\dagger \phi_{\mathbf{i}\sigma+}) \\
\tau_{\mathbf{i}}^y &= \frac{i}{2}\sum_\sigma (\phi_{\mathbf{i}\sigma-}^\dagger \phi_{\mathbf{i}\sigma+} - \phi_{\mathbf{i}\sigma+}^\dagger \phi_{\mathbf{i}\sigma-}) \\
\tau_{\mathbf{i}}^z &= \frac{1}{2}\sum_\sigma (\phi_{\mathbf{i}\sigma+}^\dagger \phi_{\mathbf{i}\sigma+} - \phi_{\mathbf{i}\sigma-}^\dagger \phi_{\mathbf{i}\sigma-}). \quad (4)
\end{aligned}$$

Since there is one quasi-particle per block B, the pseudospin has two possible flavors; i.e., $\tau$ is an $S = 1/2$ pseudospin variable. Notice that the connectors do not have any role in this low energy subspace, and the blocks B (triangles) can be replaced by effective sites containing one quasi-particle with a spin and a chiral (pseudo-spin) degree of freedom. When $V_3$ is included to first order ($V_3 \ll U, |t|$), the chiral degeneracy is removed because the inter-site Coulomb repulsion induces an $xy$ like interaction between the pseudo-spin variables. The effective model Hamiltonian is:

$$H_{eff}^{1D} = J \sum_{\mathbf{i}} (\tau_{\mathbf{i}}^x \tau_{\mathbf{i+1}}^x + \tau_{\mathbf{i}}^y \tau_{\mathbf{i+1}}^y - \frac{1}{2}), \quad (5)$$

where $J = \frac{1}{3}V_3$, and $V_3$ is the Coulomb repulsion to third nearest neighbors. Therefore, the inter-site Coulomb repulsion $V_3$ provides a microscopic mechanism for chiral ordering. The one dimensional $XY$ model [5] can be exactly solved and the ground state has critical magnetic correlations for the $x$ and $y$ spin components [12].

From the symmetry point of view, $V_3$ removes the chiral $SU(2)$ local gauge symmetry of $H(V_3 = 0)$ restricted to the lowest energy subspace and leaves a global $U(1)$ chiral symmetry generated by the total $\tau^z = \sum_{bfi} \tau_{\mathbf{i}}^z$.

If $V_1 \ll t, U$, the exact solution of the low energy spectrum of $H(V_3 = 0)$ can be extended to the region $\rho \leq \frac{1}{4}$. For $\frac{1}{8} < \rho \leq \frac{1}{4}$, there are some triangles which are doubly occupied in a chiral singlet (spin triplet) state: $\phi_{\mathbf{i}\sigma+}^\dagger \phi_{\mathbf{i}\sigma-}^\dagger |0\rangle$. When $V_3$ is included to first order, the effective Hamiltonian for the region $\frac{1}{8} < \rho \leq \frac{1}{4}$ turns to be an $XY$ model in a depleted lattice where the vacancies correspond to the triangles which are double occupied.

*Two Dimensional Lattice.* The two dimensional (2D) structure shown in Fig. 1b is a decorated Kagomè lattice. The unit cell contains nine sites (see Fig. 2c): two blocks B and three connectors C. This means that $H_0$ gives rise to nine different bands. Like in the previous case, the local chiral states created by $\phi_{\mathbf{i}\sigma+}^\dagger$ and $\phi_{\mathbf{i}\sigma-}^\dagger$ are quasi-particle operators of $H_0$ because they do not hybridize with the connectors. Since each unit cell contains two B blocks, there are four flat bands with energy $E_{1,2}^\pm(\mathbf{k}) = -t$. Two of these bands have positive (negative) chirality and $\phi_{\mathbf{i}\sigma+}^\dagger$ ($\phi_{\mathbf{i}\sigma-}^\dagger$) creates the corresponding Wannier states. The other five bands are generated by the hybridization between the Wannier states created by $\phi_{\mathbf{i}\sigma 0}^\dagger$ and the connectors. Within the other five bands, there is one which is also flat at zero energy and the dispersion relations of other four bands are given by the expression: $t \pm \sqrt{t^2 + 3t'^2[3 \pm \sqrt{3 + \gamma(\mathbf{k}) + 2\cos(k_x - k_y)}]}$ with $\gamma(\mathbf{k}) = 2(\cos k_x + \cos k_y)$. If $t > 0$ and $t \gg |t'|$, the energy of the four degenerate flat bands is again much lower than the energy of the other bands. In this situation, the lowest energy subspace of $H(V_3 = 0)$ can be exactly solved for any concentration $\rho \leq \frac{1}{9}$. The ground states of $H(V_3 = 0)$ for a given density $\rho \leq \frac{1}{9}$ again are those expressed in Eq. [3]. For this case, the block indices $\mathbf{i}_\nu$ denote the triangles in the 2D lattice and again they are all different. In particular, for $\rho = \frac{1}{9}$, there is one $\phi$ quasi-particle per block and the remaining degeneracy comes from the two possible chiralities and spin orientations. The chiral degeneracy is removed when the inter-site Coulomb repulsion, $V_3$, is included. We use again the relations given by Eqs. [4] to describe the chiral degree of freedom with a pseudospin variable. In this case, to derive the effective model it is convenient to make the transformation $\tau_{\mathbf{j}}^z \to -\tau_{\mathbf{j}}^z$, $\tau_{\mathbf{j}}^+ \to \tau_{\mathbf{j}}^-$, and $\tau_{\mathbf{j}}^- \to \tau_{\mathbf{j}}^+$, in one of the two sub-lattices of the hexagonal lattice of blocks B. After doing this transformation, the effective model is again an $S = 1/2$ $xy$ Hamiltonian on a hexagonal lattice:

$$H_{eff}^{2D} = J \sum_{\langle \mathbf{i},\mathbf{j} \rangle} (\tau_{\mathbf{i}}^x \tau_{\mathbf{j}}^x + \tau_{\mathbf{i}}^y \tau_{\mathbf{j}}^y - \frac{1}{2}), \quad (6)$$

where $J = \frac{2}{9}V_3$ and $\langle \mathbf{i}, \mathbf{j} \rangle$ indicates that $\mathbf{i}$ and $\mathbf{j}$ are n.n. triangles. In other words, the Coulomb repulsion $V_3$ induces an effective interaction between the local currents which leads to an ordered state. In the ordered state the currents are pointing in any direction contained in the $xy$ plane (the order parameter in the pseudospin notation is the transverse staggered magnetization).

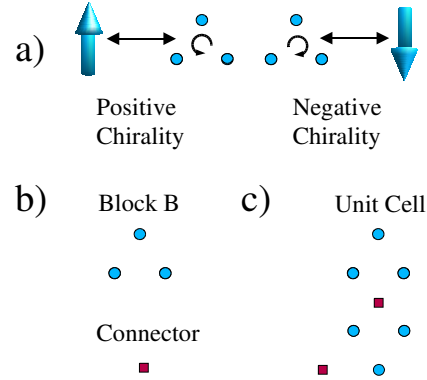

FIG. 2: a)Two possible chiral states that can occur on each block B and the connection between the chiral and the spin language. b) The block B (circles) and the connector (square) of the lattices shown in Fig.1. c) Unit cell of the decorated Kagomè lattice shown in Fig.1b.

In the same way as for the 1D case, this solution can be extended to the region $\rho \leq \frac{2}{9}$ if $V_1 \ll t, U$ and the effective model for $\frac{1}{9} \leq \rho \leq \frac{2}{9}$ is again an $XY$ Hamiltonian on a depleted lattice.

We can induce from the previous examples which is the guiding principle that provides exact solutions for the low energy spectrum of $H$ and encompasses a novel ordering. The geometrical frustration of our family of lattices gives rise to a competition between two oposite tendencies: I) Charge localization on each block B induced by a positive nearest neighbor hopping $t$ that stablizes a local orbital (see Eq.[2]) which is not hybridized with any other site on the lattice, and II) Charge delocalization favoured by the next nearest neighbor hopping $t'$. Since in general $t \gg |t'|$, the first tendency wins and a new symmetry emerges in the low energy spectrum of $H$: the local $U(1)$ gauge symmetry associated twith the conservation of the charge on each block B. The Coulomb interactions $U$ and $V_1$ prevent the double occupancy of each block B. Therefore, there is also a spin 1/2 localized on each block B that enlarges the local gauge symmetry to $U(1) \times SU(2)$. In addition, we can imagine different structures for the block B that impose a net chirality for the lowest energy orbital of the block and a consequent degeneracy. For instance, if B is a ring with an odd number of sites (in our previous examples this number is three) and $t$ is positive, there are two lowest energy orbitals with opposite chiralities [14]. This chiral internal degree of freedom increases the local gauge symmetry to $U(1) \times SU(2) \times SU(2)$.

The inclusion of the inter-site Coulomb interaction $V_3$ removes the chiral $SU(2)$ local gauge symmetry by the generation of an effective $xy$ interaction between the chiral degrees of freedom. The remaining symmetry is the global $U(1)$ group of rotations along the $z$ axis in the pseudspin space. In dimension larger than one, the $xy$ interaction induces an ordered chiral state in which the global $U(1)$ symmetry is spontaneously broken. Although in this paper we only considered one and two dimesional lattices for pedagogical reasons, the application of the same principle to three dimensional lattices is straightforward.

At this point, it is important to remark that the $U(1)$ group generated by $\tau^z = \sum_\mathbf{i} \tau^z_\mathbf{i}$ is not a symmetry of $H$. It is only a symmetry when $H$ is restricted to the low energy subspace described above. In this sense, the global $U(1)$ symmetry emerges at low energies and it is spontaneously broken at an even lower energy scale $T_c$.

In summary we have presented a new family of solvable Hubbard type models on specific lattices. The particular connectivity of these lattices is the key to generate gauge and global transformations which are symmetries for the restricted action of the Hamiltonian on an invariant subspace. If the invariant subspace corresponds to the low energy spectrum of the model, we can say that these symmetries *emerge* at low energies. In other words, the exact low energy theory has more symmetries than the original model. Using this important property we have demonstrated that the ground state of a Hubbard model on a decorated Kagomè lattice contains local currents which exhibit $xy$-like long range ordering.

*Acknowledgements.* This work was sponsored by the US DOE. BSS acknowledges support from the Indo French grant IFCPAR/2404-1. We thank B. Kumar and J. E. Gubernatis for helpful comments.

‡ Also at JNCASR, Bangalore, INDIA.